\documentclass[prb,aps,floats,twocolumn]{revtex4}
\usepackage{tabularx,graphicx}
\usepackage{epsfig}
\usepackage{amsmath}
\usepackage{amsfonts}
\usepackage{bbm}
\usepackage{graphicx}
\begin{document}

\title{Topological protection from exceptional points in 
Weyl and nodal line semimetals}
\author{J. Gonz\'alez  and R.A. Molina\\}
\affiliation{Instituto de Estructura de la Materia,
        Consejo Superior de Investigaciones Cient\'{\i}ficas, Serrano 123,
        28006 Madrid, Spain}

\date{\today}

\begin{abstract}
We investigate the topological protection of surface states in Weyl and nodal-line semimetals by characterizing them as evanescent states when the band structure is extended to complex momenta. We find in this way a sequence of exceptional points ---that is, branch points with zero energy in the complex spectrum--- allowing us to identify the set of surface states with complex momentum signaling the decay into the 3D semimetal. 
From this point of view, Weyl and nodal-line semimetals can be classified in two types depending on the way surface states decay.
Type A semimetals have surface states with smaller penetration length and oscillating decay while type B semimetals have longer simple exponential decays. The difference between both types reflects in the way the branch cuts in the spectrum accommodate in the complex plane. 
The stability of the surface states stems in this approach from the complex structure that develops around the exceptional points, with a topological protection which is based on the fact that the branch cuts cannot be closed by small perturbations. We check this property when nodal-line semimetals are placed under circularly polarized light, where we observe that the exceptional points survive the effect of such a perturbation, though appropriate boundary conditions for zero-energy surface states cannot be satisfied in general due to the breakdown of time-reversal invariance by the radiation field.
\end{abstract}

\maketitle

\date{\today}

\vspace{2cm}

\section{Introduction}

Topological materials have become a primary target of research in the last few years due to their exceptional properties with high potential for applications \cite{top1,top2}. The bulk-edge correspondence and the topological protection of edge states is on the basis of the interesting properties of systems with topologically non-trivial band structure. In topological insulators, the quantity controlling the topological protection is the inverted gap between bands arising from the crystal structure of the material. A perturbation must be strong enough to close the gap in order to destroy the
edge states of the material.

In band theory, a semimetal is usually defined as a material with no gap in the band structure but with a very small overlap between the valence and the conduction band, resulting in a small density of states around the Fermi energy. A typical example is bismuth which has intermediate properties between an insulator and a metal \cite{Burns_book}. In topological band theory, however, it is customary to work with a more strict definition by which a semimetal is a material with no gap but also no Fermi surface, so the conduction and valence bands touch only at isolated points. The semimetal so defined is really a distinct phase intermediate between insulators and metals, the most famous example being graphene \cite{novo}. Although the Dirac nodes in the 2D Brillouin zone of graphene are not topologically protected, in three dimensions we have the last additions to the family of topological materials, which include Dirac and Weyl semimetals with isolated Dirac or Weyl nodes in the band structure \cite{liu,neupane,borisenko,taas1}, and the nodal-line semimetals with a continuous line of nodes in the Brillouin zone \cite{Burkov16}.


The surface states corresponding to these topological semimetals lie on constant energy contours which do
not form in general closed curves. In the case of the Weyl and Dirac semimetals, the surface states lead to the celebrated
Fermi arcs joining the projection of the nodes onto the given surface. For nodal-line semimetals, the surface states form
the so-called drumhead within the nodal circle \cite{Phillips14,Chiu14,Burkov16}. In the Weyl semimetals, 
such states are topologically protected and Chern numbers can be defined in the planes lying between 
Weyl nodes \cite{Wan11,Yang11}. The physical quantity measuring this protection is the separation of the nodes
in momentum space. The situation is less clear for Dirac semimetals in which a simple application of
bulk-surface correspondence does not provide with an answer to whether or not the surface states are 
topologically protected. Recent works have discussed the stability of the
surface states in 3D semimetals \cite{Kargarian16,gorbar,serguei}. In this regard, an alternative description of the topological protection of surface 
states in 3D topological semimetals is very desirable. Here, we provide such a description by extending 
the band structure in momentum space to complex momenta. 

The convenience of taking complex values of the momentum is motivated by the search of evanescent 
states in the 3D semimetals. When the momentum is promoted to
a complex quantity, the Hamiltonian of the system becomes non-Hermitian, but it is still
possible to find exceptional points in the spectrum, that is, branch points where the imaginary part
of the eigenvalue vanishes. This allows one to identify the set of evanescent states with complex 
momentum that signals the decay into the 3D semimetal. Such states are endowed with topological protection,
which arises naturally from the fact they are attached to branch cuts that cannot be removed under small
perturbations. This description based on the identification between evanescent states and exceptional points was pioneered in a previous work describing surface states in Weyl semimetals in the presence of circularly polarized light \cite{Gonzalez16}. 
Recently, rings of exceptional points have been also found in dissipative systems which are based on Weyl semimetals \cite{Xu}.

When looking for evanescent states in the spectrum of the 3D semimetals, we have found that these 
can be classified in two different groups, depending on the pattern of the branch cuts in the plane of
complex momentum. Thus, there is a class of 3D semimetals, that we denote as type A, where most
part of the branch cuts run in parallel crossing the real axis. The corresponding exceptional points form 
then quartets belonging to the same branch of the spectrum, with each member of a quartet 
in a different quadrant of the complex plane. In the other class, that we denote as type B, all the 
branch cuts can be disposed instead along the imaginary axis, lacking the nontrivial
realization of symmetry found in the type A class. From the physical point of view, this introduces also 
an important difference between the two classes, as the quartets found in the type A semimetals provide a 
higher degree of topological protection, quantified in terms of a larger length of the branch cuts 
and a much smaller penetration length of the evanescent states in the 3D semimetal. 

Note that this classification has a 
different origin than the classification into type I and type II Weyl semimetals, that depends on whether the density of
states vanishes at the nodal points or has some extra contributions due to the tilting of the Weyl nodes \cite{Soluyanov15,Xu3}.
Although we have only explored a model describing a type I Weyl semimetal in this work, we expect our classification to be
supplementary to the classification into type I and type II Weyl semimetals, that is, we expect to be type A semimetals of type I and type II and type B semimetals also of both type I and II.  
  
The stability of the surface states stems in our approach from the complex structure that develops 
for complex momentum, where different bands can be seen as different branches of the Riemann surface 
giving the spectrum.  
In this framework, a pair of exceptional points and the respective evanescent states can be only annihilated 
by merging the branch points, provided they lie in the same branch of the spectrum. This picture allows
us to establish a connection with the usual account of the topological protection of 
surface states, which requires making a 2D projection of the 3D band structure to open a gap in the 
spectrum. In our approach a gap also exists, but this is now seen as the separation opened between 
different branches along a branch cut in the complex spectrum. Such a gap can be closed only as long as 
the exceptional points at the two ends of the branch cut are made to coalesce, which provides an 
alternative understanding of the topological protection in the plane of complex momentum.

In this paper we apply the complex structure developed around the exceptional points to investigate
the stability of the surface states in 3D Weyl and nodal-line semimetals. We are 
going to see that this approach provides a very robust picture of the evanescent states in these
systems. This will be checked in particular in the case of the nodal-line semimetals under circularly polarized light, which is an instructive example since that may not be in general 
a small perturbation of the semimetal. We will see that the exceptional points survive indeed the effect of the 
electromagnetic field, though appropriate boundary conditions for surface states cannot be satisfied in general due to the breakdown of time-reversal invariance by the radiation field.




The paper is 
organized as follows. In Sec. \ref{sec:weyl}, we consider Weyl semimetals and their
surface states on the light of our approach to look for evanescent states with complex momentum. Then, in Sec. \ref{sec:nodalline} we consider nodal-line semimetals, their bulk and their 
surface states in cylindrical coordinates which are well-suited to the problem. In Sec. 
\ref{sec:nodalfloquet}, we use the same approach to study the behavior of evanescent 
states in the case of a periodic perturbation of the nodal-line semimetal
by circularly polarized light. We finish with some conclusions of our study in Sec. \ref{sec:conclusions}.

\section{Weyl semimetal}
\label{sec:weyl}

We consider a simple model for a Weyl semimetal with Hamiltonian
\begin{equation}
H_{\rm w} = (m_0 + m_1 \boldsymbol{\nabla}^2 ) \sigma_z 
      - iv \partial_z \sigma_x - iv \partial_y \sigma_y.
\label{hw}
\end{equation}
The energy-momentum dispersion as a function of the 3D momentum ${\bf k}$
is given then by 
\begin{equation}
\varepsilon = \pm \sqrt{(m_0 - m_1 \mathbf{k}^2)^2 + v^2 k_y^2 + v^2 k_z^2 }.
\end{equation}
It turns out that the valence and conduction bands touch at Weyl points located 
in the line $k_y = k_z = 0$ with
\begin{equation}
k_x = \pm \sqrt{\frac{m_0}{m_1}}.
\end{equation}

In this model we may look for surface states characterized by wavefunctions 
decaying for instance in the $z$ direction as
\begin{equation}
\phi (x,y,z) \sim  e^{ik_z z} e^{-\alpha z} f(x,y),
\end{equation}
with $\alpha > 0$ corresponding to the inverse of the penetration length.
The action of the Hamiltonian becomes particularly simple if we concentrate 
on the set of states 
\begin{eqnarray}
\phi_{k_x,k_z,+} (x,y,z) & = & 
                   e^{ik_x x} e^{ik_z z} e^{-\alpha z} | + \rangle  
                                                \label{b1}        \\
\phi_{k_x,k_z,-} (x,y,z) & = & 
                    e^{ik_x x} e^{ik_z z} e^{-\alpha z} | - \rangle
\label{b2}
\end{eqnarray}
with the spinor part corresponding to the eigenvectors of $\sigma_y $
\begin{equation}
| + \rangle =  \left(\begin{array}{c} 1  \\  i  \end{array} \right)
\;\;\;\;\;  ,  \;\;\;\;\;
| - \rangle =  \left(\begin{array}{c} 1  \\  -i  \end{array} \right).
\label{sp}
\end{equation}
We get in this way
\begin{widetext}
\begin{eqnarray}
H_{\rm w} \phi_{k_x,k_z,+} &=&   
    \left( m_0 - m_1 (k_x^2 + k_z^2 - \alpha^2 + 2ik_z \alpha ) 
             + iv (k_z + i\alpha )  \right) \phi_{k_x,k_z,-} \label{zm1} \\
H_{\rm w} \phi_{k_x,k_z,-} &=&   
    \left( m_0 - m_1 (k_x^2 + k_z^2 - \alpha^2 + 2ik_z \alpha ) 
               - iv (k_z + i\alpha )  \right) \phi_{k_x,k_z,+}. 
\label{zm2}
\end{eqnarray}
\end{widetext}

Then we can identify a collection of zero-energy modes in 
the set of states $\{ \phi_{k_x,k_z,+} \}$ 
by canceling out the 
right-hand-side of Eq. (\ref{zm1}) (assuming that $m_0 > 0, m_1 > 0, v > 0$). 
This leads to two types of solutions, either
\begin{eqnarray}
\alpha  & = &  \frac{v}{2 m_1}          \label{c1}   \\
 k_z  & = &  \pm \sqrt{ \frac{m_0}{m_1} - k_x^2 - \alpha^2} \label{c2}
\end{eqnarray}
or
\begin{eqnarray}
\alpha &=& \frac{v\pm\sqrt{v^2-4m_1(m_0-m_1k_x^2)}}{2m_1} \label{c3} \\
k_z &=& 0 \label{c4}
\end{eqnarray}
For a given $k_x$, two independent states with $\alpha > 0$ exist as long as
$k_x^2 \leq m_0 /m_1$. We note that the actual wave function corresponding to 
a surface state must be a linear combination of the solutions with $+$ and $-$ 
signs (in either (\ref{c1})-(\ref{c2}) or (\ref{c3})-(\ref{c4})) in order to 
fulfill appropriate boundary conditions (for example in a semi-infinite plane,  
making the wave function to vanish at $z = 0$). This collection of evanescent 
states maps therefore the celebrated Fermi arcs with $|k_x| \leq \sqrt{m_0 /m_1}$, 
which join the projection of the Weyl points onto a given surface of the semimetal.

The solution corresponding to Eqs. (\ref{c1})-(\ref{c2}) is valid when 
$4m_1m_0 > v^2$. In this case, there is however a portion of the Fermi arcs,
closer to the endpoints, where the states change to the form given by 
Eqs. (\ref{c3})-(\ref{c4}). In the regime $4m_1m_0 < v^2$, all the states in the 
Fermi arcs correspond instead to this latter representation of the evanescent
eigenstates. In what follows, we will see that there is indeed a clear 
distinction between two different types of surface states, allowing us to 
discern two different regimes of Weyl semimetals that we denote as type A 
(for $4m_1m_0 > v^2$) and type B (for $4m_1m_0 < v^2$).

\begin{figure}
\includegraphics[width=0.4\textwidth]{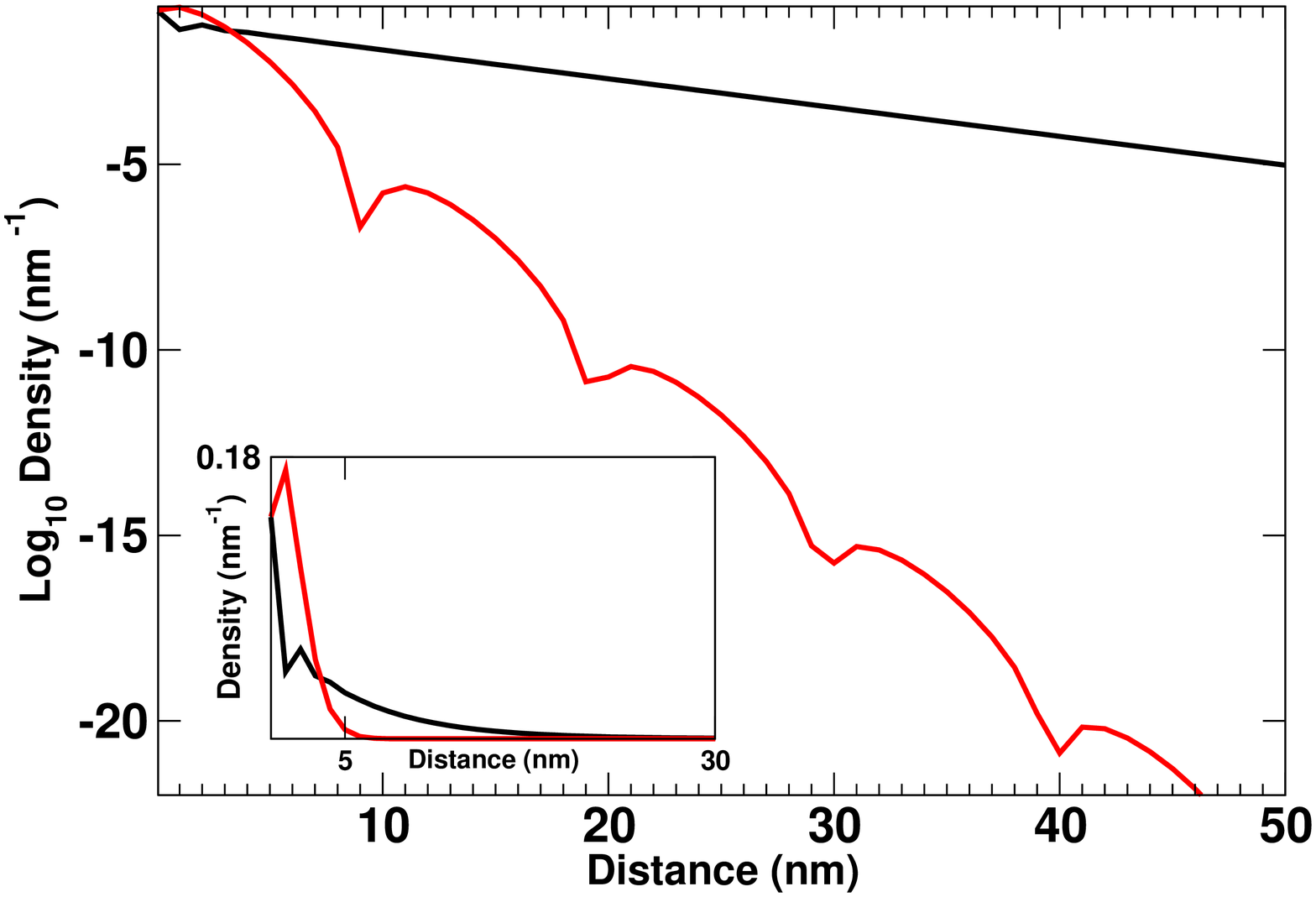}
\caption{Probability density in logarithmic scale across the $z$ direction of surface states in the Fermi arcs with $k_x=0$. We compare two different cases, one with some oscillatory dependence in the $z$ direction corresponding to a type A Weyl semimetal (red line, $m_0=0.35$ eV, $m_1=1.0$ eV  nm$^2$, $v=1.0$ eV nm) and another with pure exponential decay corresponding to a type B Weyl semimetal (black line, $m_0=0.35$ eV, $m_1=1.0$ eV nm$^2$, $v=4.0$ eV nm). In the inset the same figure is shown in normal scale. \label{fig:fermiarcs} }
\end{figure}

In Fig. \ref{fig:fermiarcs} we show numerical calculations for an equivalent tight-binding Hamiltonian in a slab of width $W=1000$ nm in the $z$ direction, for two different examples pertaining to type A and type B semimetals. The appearance of the band structure is very similar in the two regimes, with the Fermi arcs joining the projection of the Weyl nodes. However, 
in the type B regime ($4m_1m_0<v^2$) the Fermi arc states have a penetration length
(1/$\alpha$) that increases with the value of $|k_x|$ until it diverges at the projection of the Weyl nodes.
In the type A regime ($4m_1m_0>v^2$), the states with $k_x^2 < m_0/m_1-v^2/4m_1^2$ have a very small penetration length $2m_1/v$ which would be typically of the order of a few unit cells in real materials. However, the probability density oscillates as it decays. For larger values of $k_x^2$, the penetration length grows until it diverges again at the projection of the Weyl nodes, and the wave function for $z$ is just a decaying exponential without oscillations.

Within this approach, we can also analyze the role that the evanescent states
play in the spectrum of the Hamiltonian $H_{\rm w}$, as this becomes a 
non-Hermitian operator acting on the basis (\ref{b1})-(\ref{b2}). In the 
subspace spanned by this set of states, the eigenvalues of $H_{\rm w}$ turn 
out to be according to (\ref{zm1})-(\ref{zm2})
\begin{equation}
\lambda = \pm \sqrt{ \left(m_0 - m_1 (k_x^2 + (k_z + i \alpha )^2) \right)^2
                     + v^2   (k_z + i \alpha )^2  }
\label{lam}
\end{equation}
It can be seen that the particular values satisfying the zero-mode conditions
(\ref{c1})-(\ref{c2}) and (\ref{c3})-(\ref{c4}) correspond to branch points in the complex spectrum of 
$H_{\rm w}$, both for $4m_0m_1>v^2$ and $4m_0m_1<v^2$. 
We find therefore that the evanescent states we have identified correspond
to so-called exceptional points \cite{berry,heiss} in the spectrum of $H_{\rm w}$, 
when this is mapped as a function of the complex momentum $k_z + i \alpha $. 
This highlights that there is a complex structure behind the surface states of 
the Weyl semimetals, which has important implications for their stability.

\begin{figure}[t]
\mbox{}
\\
\begin{center}
\mbox{\epsfxsize 8.1cm \epsfbox{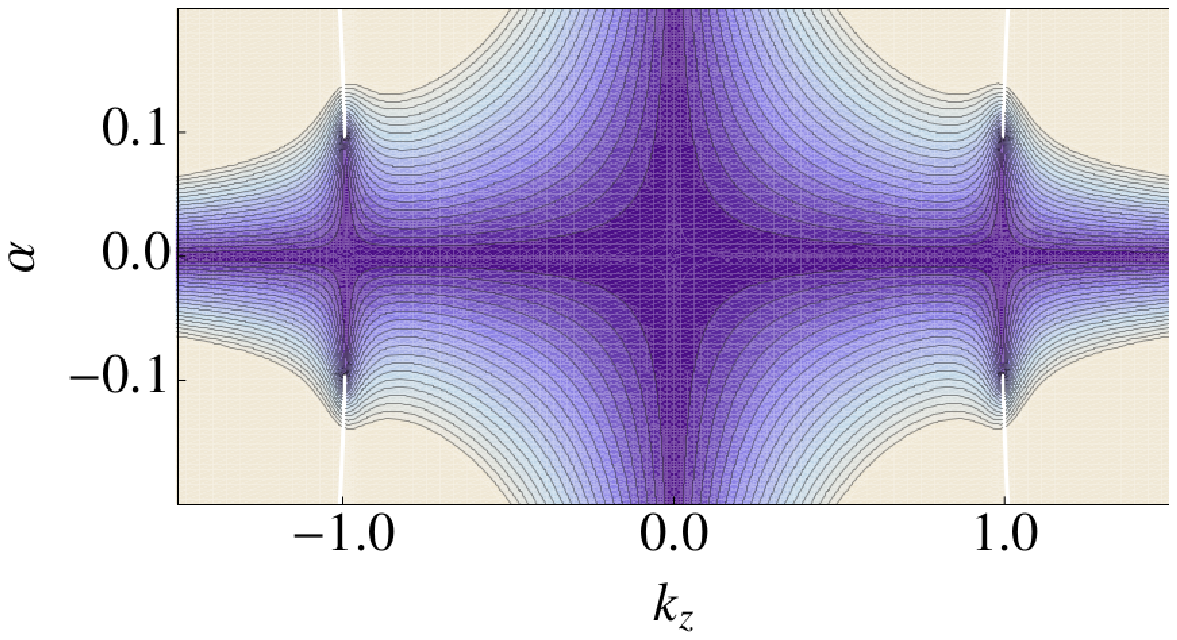}}
\end{center}
\begin{center}
(a)
\end{center}
\begin{center}
\mbox{\epsfxsize 8.1cm \epsfbox{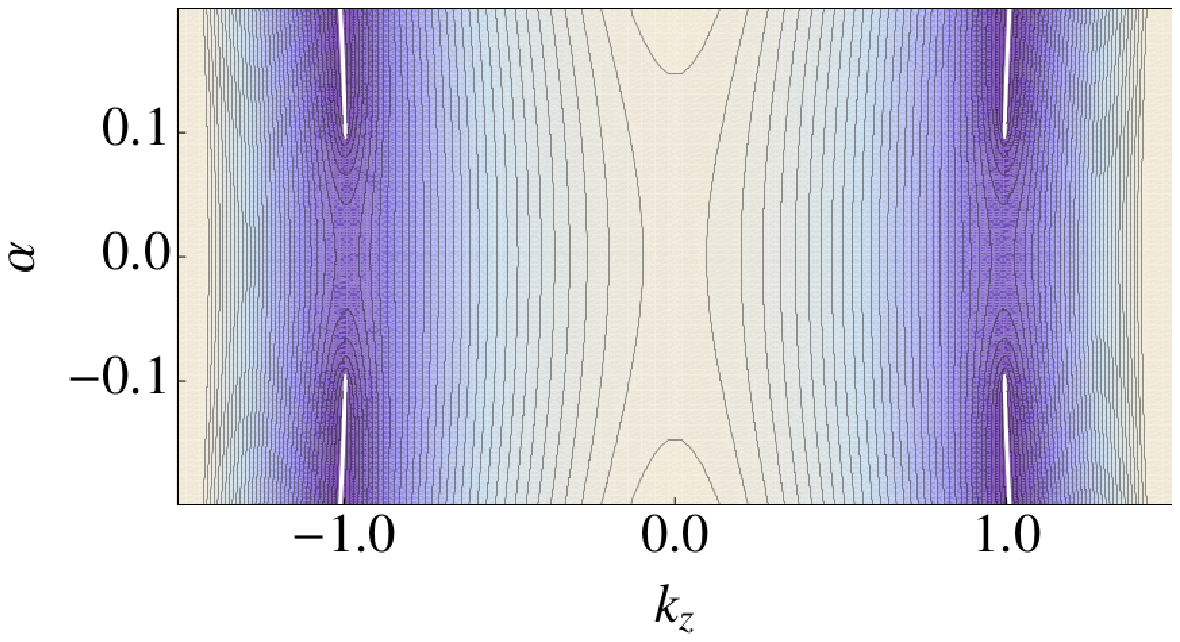}}
\end{center}
\begin{center}
(b)
\end{center}
\caption{Contour plots of the absolute value of the imaginary part (a) 
and the real part (b) of the 
eigenvalue in (\ref{lam}) for $k_x = 0$ and $m_0 = 1.0$ eV, $m_1/a^2 = 1.0$ eV, 
$v/a = 0.2$ eV, $a$ being a microscopic length scale in the model. 
$k_z$ and $\alpha $ are measured in units of $a^{-1}$.
Lighter colors correspond to increasing energy values and dark colors 
represent regions close to zero.}
\label{one}
\end{figure}

\begin{figure}[t]
\mbox{}
\\
\begin{center}
\mbox{\epsfxsize 8.1cm \epsfbox{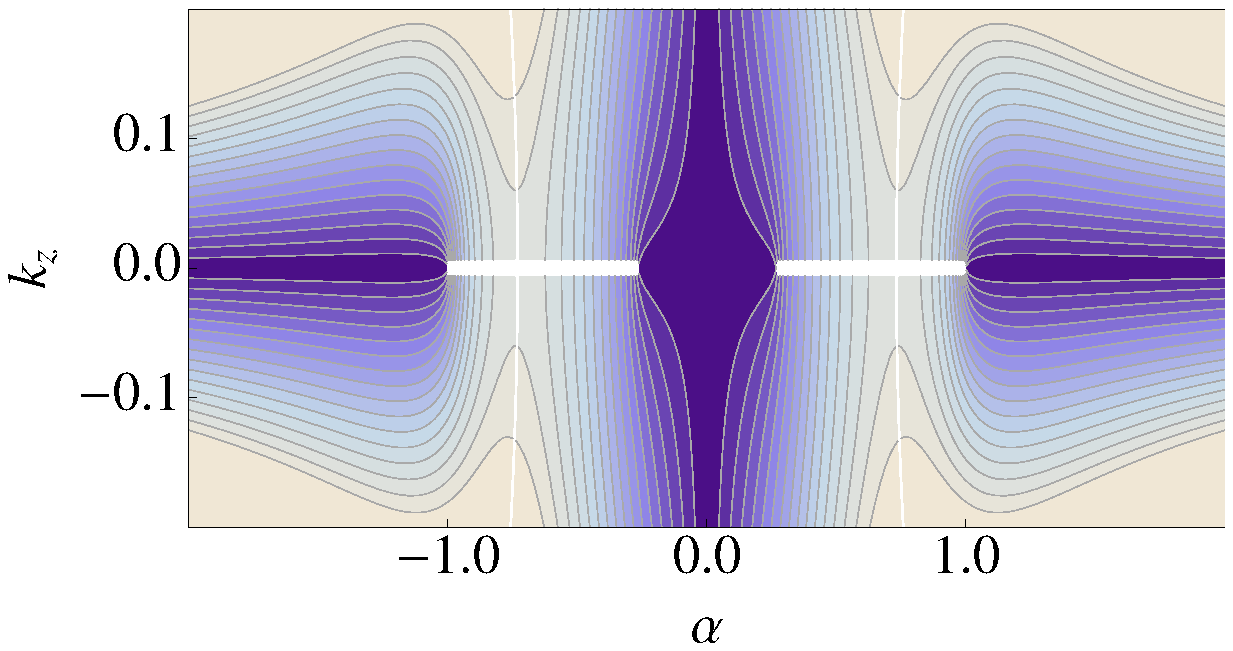}}
\end{center}
\begin{center}
(a)
\end{center}
\begin{center}
\mbox{\epsfxsize 8.1cm \epsfbox{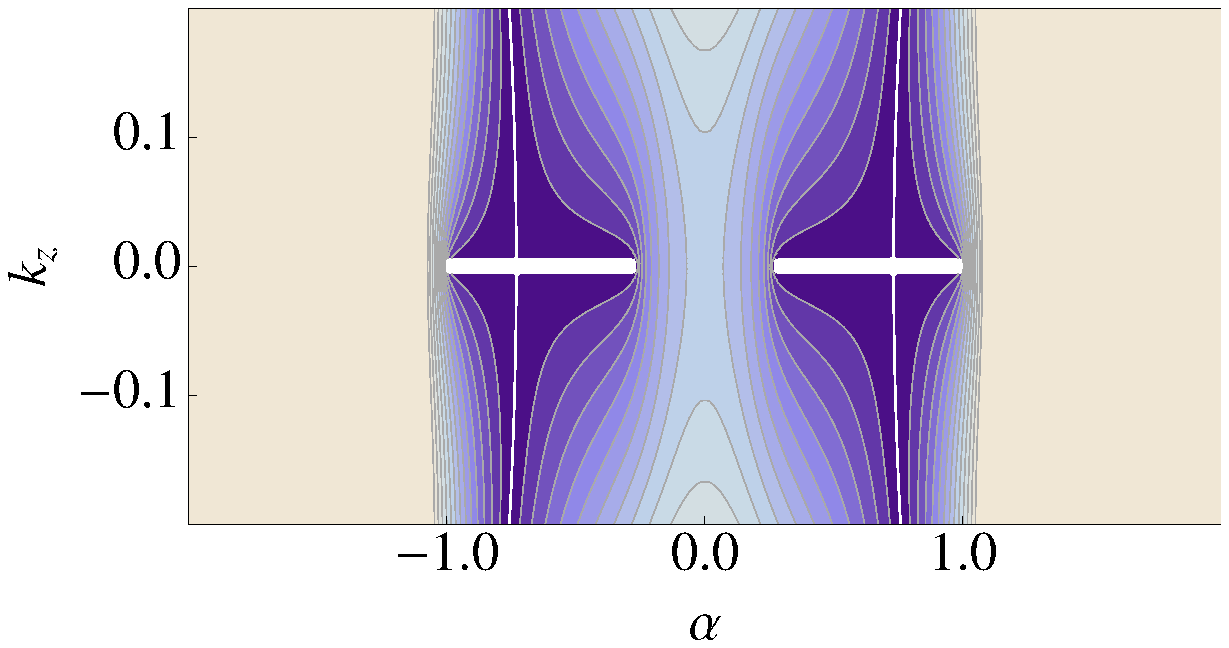}}
\end{center}
\begin{center}
(b)
\end{center}
\caption{Contour plots of the absolute value of the imaginary part (a) 
and the real part (b) of the 
eigenvalue in (\ref{lam}) for $k_x = 0$ and $m_0 = 0.5$ eV, $m_1/a^2 = 1.9$ eV, 
$v/a = 2.4$ eV, $a$ being a microscopic length scale in the model.
$k_z$ and $\alpha $ are measured in units of $a^{-1}$.
Lighter colors correspond to increasing energy values and dark colors 
represent regions close to zero.}
\label{oneb}
\end{figure}

The structure of the branch cuts in the complex plane $k_z + i\alpha $ is
different however, depending on whether we consider the type A regime
($4m_0m_1>v^2)$ or the type B regime ($4m_0m_1<v^2$) of the Weyl semimetal.
In the first case, the states in the Fermi arcs with $k_x^2 < m_0/m_1-v^2/4m_1^2$ 
correspond to branch points that are away from the imaginary axis, as represented 
in Fig. \ref{one}, where it can be observed the branch cuts that develop from 
the location of the zero-energy modes. The branch cuts fall eventually into 
the imaginary axis for $k_x^2 > m_0/m_1-v^2/4m_1$, where we know that the 
evanescent states must be in accordance with (\ref{c3})-(\ref{c4}). In the type
B regime, however, the branch cuts are found in the imaginary axis for all the
states in the Fermi arcs, with a typical structure represented in Fig. \ref{oneb}.

The different structure of the branch cuts in the type A and type B regimes
of the Weyl semimetal can be actually ascribed to a different realization of the
symmetries of the Hamiltonian (\ref{hw}). This is in particular invariant 
under the spatial inversion, which can be represented as an 
operator $I$ with an action on the states $\phi (x,y,z)$ given by 
\begin{equation}
I: \; \phi (x,y,z) \rightarrow  \sigma_z \: \phi (-x,-y,-z)
\label{si}
\end{equation}
The Hamiltonian (\ref{hw}) has moreover an enlarged symmetry when the dynamics 
is constrained to states that do not depend on the $y$ coordinate, as in the 
above discussion. Then $H_{\rm w}$ becomes invariant under a transformation
$T$ that acts like time-reversal invariance, given in terms of the operation
of complex conjugation $K$ as
\begin{equation}
T: \; \phi (x,y,z) \rightarrow  K \sigma_z \: \phi (x,y,z)
\label{tr}
\end{equation}
Thus, in the structure represented in Fig. \ref{one} for the 
type A regime of the Weyl semimetal, the two branch points related by the 
inversion of the complex momentum $k_z + i \alpha $ correspond to states that 
are mapped onto each other by the action of $I$. On the other hand, the 
transformation $T$ is realized in the complex plane as the 
inversion $k_z \rightarrow -k_z$. This accounts for the fact that the 
same state $\phi (x,y,z)$ (with $y = {\rm const.}$) is found at the two 
branch points related by such an inversion of the momentum $k_z$.

The fact that one finds the same evanescent state at branch points with 
momenta $k_z$ and $-k_z$ is important from the point of view
of the stability of the surface states. As already mentioned, two independent 
states $\phi (x,y,z)$ are needed in order to build a surface state that may
vanish on the boundary of the system, say for instance on the plane $z = 0$. This 
requirement can be satisfied in the type A regime of the Weyl semimetal from 
the symmetry enforced by the $T$ operation. Then, the surface states 
with $k_z \neq 0$ turn out to be protected due to the particular structure of 
the branch cuts, which cannot be undone unless the branch points are made to 
coalesce in pairs. 
Such kind of topological protection holds also but to a lesser degree in the 
type B regime of the Weyl semimetal.
In that case, the evanescent states
correspond to branch points with more variable separation along
the imaginary axis, which can merge therefore under weaker perturbations. 

We find that the exceptional points endow in general the surface 
states with topological protection, as the nontrivial topology of the branch 
cuts cannot be modified with small perturbations. When the exceptional points
realize the $I$ and $T$ symmetry operations, that protection can be also 
quantified in terms of the distance separating each pair in a quartet of 
exceptional points. Alternatively, this protection can be also understood 
from the existence of a gap that opens up along each branch cut, as observed
in Figs. \ref{one} and \ref{oneb}, though here we realize that this is just an
effect of a more involved picture relying on the full complex structure 
unveiled in the $(k_z, \alpha )$ plane.

\section{Nodal line semimetal}
\label{sec:nodalline}

\subsection{Bulk states}
\label{subsec:nodalbulk}

Our starting point is a model of nodal line semimetal with a Hamiltonian
\begin{equation}
H_{\rm NL} = (m_0 + m_1 \boldsymbol{\nabla}^2 ) \sigma_z - iv \partial_z \sigma_x
\label{ham0}
\end{equation}
In terms of the 3D momentum $\mathbf{k}$, the eigenvalues of (\ref{ham0}) are 
given by 
\begin{equation}
\varepsilon = \pm \sqrt{(m_0 - m_1 \mathbf{k}^2)^2 + v^2 k_z^2 }
\end{equation}
This model has then a line of nodes in the plane $k_z = 0$,
given by the circular set
\begin{equation}
k_x^2 + k_y^2 = \frac{m_0}{m_1}
\end{equation}

When looking for energy eigenstates, we can concentrate on modes
with well-defined angular momentum, with wavefunction $\psi $ such that in
polar coordinates $(r, \theta, z )$
\begin{equation}
\psi (r, \theta, z)  \sim  e^{i k_z z} e^{i m \theta } f(r)
\end{equation}
The spectrum can be obtained then by solving the eigenvalue problem
\begin{widetext}
\begin{equation}
\left[m_0 + m_1 \left( \frac{1}{r} \frac{\partial }{\partial r} 
  \left(r \frac{\partial }{\partial r} \right) - \frac{m^2}{r^2}  
      - k_z^2 \right)  \right] \sigma_z \psi_{k_z}^{(m)} 
      + v k_z \sigma_x \psi_{k_z}^{(m)}  =  \varepsilon  \psi_{k_z}^{(m)}
\label{ep}
\end{equation}
\end{widetext}

In the case of bulk zero-energy modes, we see from (\ref{ep}) that
they can be expressed in terms of Bessel functions $J_m$ as 
\begin{eqnarray}
\psi_{0,+}^{(m)} (r, \theta, z)  & = & 
          e^{i m \theta } J_m (k r) | u \rangle     \\
\psi_{0,-}^{(m)} (r, \theta, z)  & = & 
          e^{i m \theta } J_m (k r) | d \rangle
\end{eqnarray}
with $k = \sqrt{m_0 /m_1 }$ and the spinor part given by the eigenvectors of 
$\sigma_z $
\begin{equation}
| u \rangle =  \left(\begin{array}{c} 1  \\  0  \end{array} \right)
\;\;\;\;\;  ,  \;\;\;\;\;
| d \rangle =  \left(\begin{array}{c} 0  \\  1  \end{array} \right)
\end{equation}
It turns out that, in this particular representation, the states corresponding 
to the nodal line are labeled by the integer values $m$ of the projection
of the angular momentum along the $z$ direction.

\subsection{Surface states}
\label{subsec:nodalsurface}

We are interested in surface states that take the form of evanescent 
waves localized at the boundary of the 3D semimetal. It can be easily 
seen that there is a huge set of these states characterized by the 
evanescence in the $z$ direction, with wavefunction $\chi $ decaying as
\begin{equation}
\chi (r, \theta, z)  \sim  e^{i k_z z} e^{-\alpha z}  e^{i m \theta } f(r)
\label{dec}
\end{equation}
The zero-energy modes have to correspond in particular to solutions of the 
equation
\begin{widetext}
\begin{equation}
\left[m_0 + m_1 \left( \frac{1}{r} \frac{\partial }{\partial r} 
  \left(r \frac{\partial }{\partial r} \right) - \frac{m^2}{r^2}  
       - k_z^2 + \alpha^2 - 2i k_z \alpha \right)  \right] \sigma_z \chi_{k_z}^{(m)} 
      + v (k_z + i\alpha ) \sigma_x \chi_{k_z}^{(m)}  =  0
\label{ev2}
\end{equation}
\end{widetext}

First of all, the imaginary terms must cancel out in (\ref{ev2}). This means 
that the solutions must be necessarily proportional to the eigenvectors 
$| \pm \rangle$ of $\sigma_y$ given in (\ref{sp}).
Taking the first spinor, one obtains the constraint
\begin{equation}
- 2i m_1 k_z \alpha + i v k_z = 0
\label{cond}
\end{equation}
which leads to either 
\begin{equation}
\alpha = \frac{v}{2m_1 }
\label{alf}
\end{equation}
or
\begin{equation}
k_z=0.
\label{kz0}
\end{equation}
Similarly to the case of the Weyl semimetals, the condition (\ref{alf}) gives 
rise to evanescent states with oscillatory decay while the condition (\ref{kz0}) 
results in an exponential decay without oscillations.
If we choose otherwise the second spinor in (\ref{sp}), that changes 
the sign of the last term in the left-hand-side of (\ref{ev2}), which prevents 
the existence of evanescent states with $k_z \neq 0$ (for $v > 0$, $m_1 > 0$) 
and does not allow either to find solutions with $\alpha > 0$ when $k_z = 0$
(as we see in what follows).

Taking the value of $\alpha $ in (\ref{alf}), it turns out that the 
evanescent states are given by the solutions of Eq. (\ref{ev2})
\begin{equation}
\chi_{k_z,+}^{(m)} (r, \theta , z) = e^{i k_z z} e^{-\alpha z} 
               e^{im \theta } J_m (k_r r )  | + \rangle
\label{set1}
\end{equation}
with  
\begin{equation}
k_r = \sqrt{\tfrac{m_0}{m_1} - k_z^2 - \alpha^2 }.
\label{ktr}
\end{equation}
If instead we adopt the condition (\ref{kz0}), the corresponding values of 
$\alpha $ become
\begin{equation}
\alpha  =  \frac{v\pm\sqrt{v^2-4m_1(m_0-m_1k_r^2)}}{2m_1}
\label{alf2}
\end{equation}
which implies a penetration length $1/\alpha $ diverging at the line of nodes. 
The relation (\ref{ktr}) only makes sense if $4m_0 m_1 > v^2$, while in the 
regime $4m_0 m_1 < v^2$ all the evanescent states are found using (\ref{alf2}).
As in the case of the Weyl semimetal, this disjunctive allows us to distinguish
between two different classes of nodal line semimetals, that we denote as 
type A (for $4m_0 m_1 > v^2$) and type B (for $4m_0 m_1 < v^2$). The type A 
corresponds to the regime that is expected to hold for realistic materials 
providing examples of nodal line semimetals. From the physical point of view,
the two classes A and B can be discerned by the different penetration of the 
drumhead surface states into the material, which shows for a slab practically 
the same behavior as represented in the case of the Weyl semimetal in Fig. 
\ref{fig:fermiarcs}. 

\begin{figure}[h]
\begin{center}
\mbox{\epsfxsize 8.1cm \epsfbox{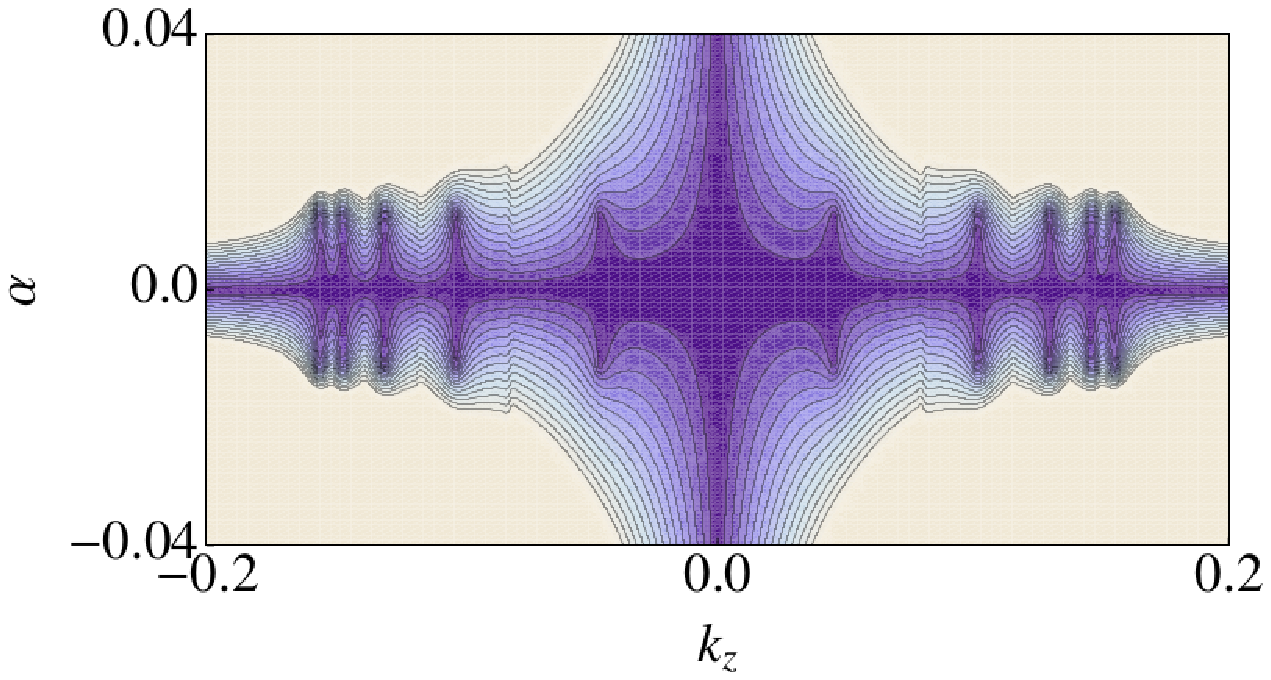}}
\end{center}
\begin{center}
(a)
\end{center}
\begin{center}
\mbox{\epsfxsize 8.1cm \epsfbox{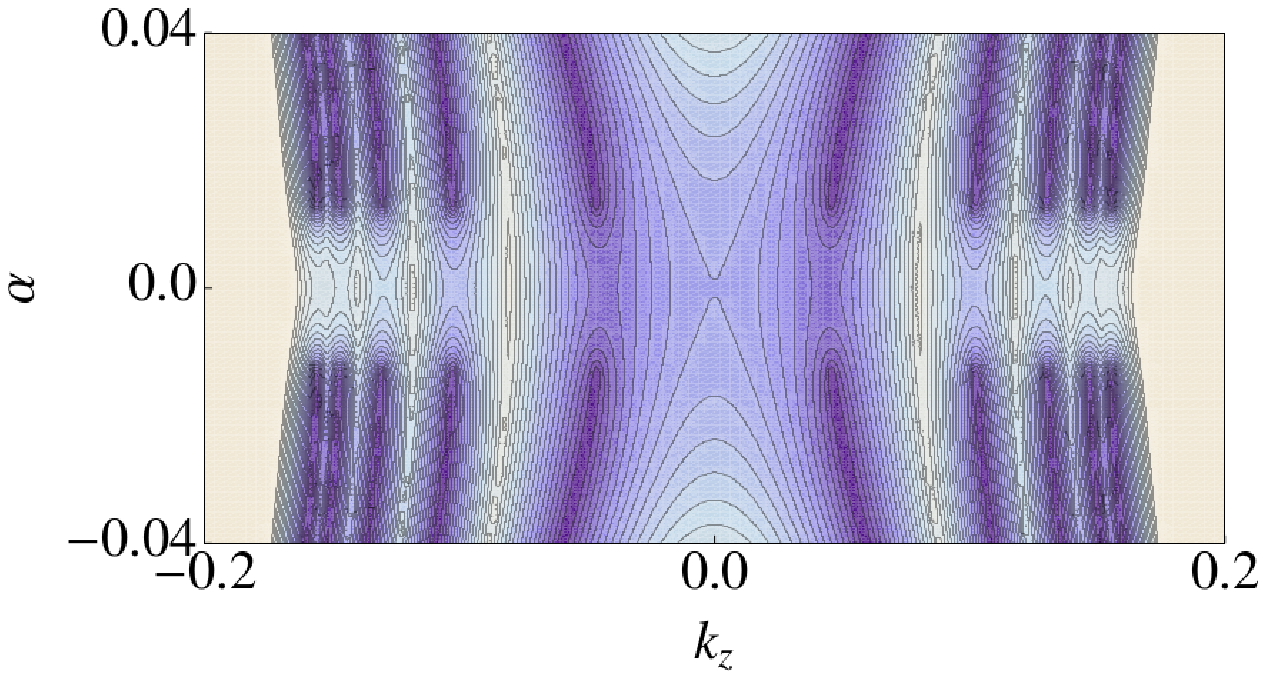}}
\end{center}
\begin{center}
(b)
\end{center}
\caption{Contour plots of the absolute value of the imaginary part (a) 
and the real part (b) of the 
lowest eigenvalue in the spectrum of $H_{\rm NL}$ for evanescent states with 
angular momentum $m = 0$ in a 
semi-infinite cylindrical geometry with radius $R = 100 a$ 
($a$ being the microscopic length scale in the model), for 
$m_0 = 0.1$ eV, $m_1/a^2 = 4.0$ eV, $v/a = 0.1$ eV. The units and the 
intensity code are the same as in Fig. \ref{one}.}
\label{two}
\end{figure}

From a formal point of view, the difference between type A and type B nodal
line semimetals lies in the distinctive complex structures that develop in the 
plane $(k_z, \alpha )$. In a type A nodal line semimetal which has for instance 
a finite circular section at the boundary $z = 0$, the values of $k_r$ become 
quantized, which turns into the consequent quantization of the momentum $k_z$. 
It can be seen that the allowed values of $k_z + i \alpha $ leading to 
evanescent states emerge then as exceptional points in the extension of the 
momentum $k_z$ to the complex plane, as represented in Fig. \ref{two}. Those 
can be characterized indeed as branch points in the spectrum of the Hamiltonian 
for complex momentum, leading to branch cuts that run down to homologous branch 
points with the reversed sign of $\alpha $.

We observe in Fig. \ref{two} that the exceptional points can be grouped 
forming quartets, which is a consequence of the invariance of the Hamiltonian
(\ref{ham0}) under the operations defined by (\ref{si}) and (\ref{tr}).
We have actually
\begin{eqnarray}
\left[H_{\rm NL}, I \right]   &  =  &  0           \\ 
   \left[H_{\rm NL} , T \right]   &  =  &  0 
\end{eqnarray}
The invariance under $T$ is responsible for the fact that each pair of evanescent 
states with opposite sign of $k_z$ may have the same wavefunction 
at $z = 0$. This symmetry is crucial in order to enforce the boundary
conditions for the surface states at the edge of the semimetal, allowing to build 
for instance linear combinations of evanescent states that vanish at $z = 0$.     
Moreover, the stability of the surface states is also guaranteed by the 
separation in the $(k_z, \alpha )$ plane between evanescent states related by
the $I$ and $T$ operations, which lend topological protection as the branch cuts
running between the respective branch points cannot be closed under small
perturbations.

\begin{figure}[t]
\mbox{}
\\
\begin{center}
\mbox{\epsfxsize 7.0cm \epsfbox{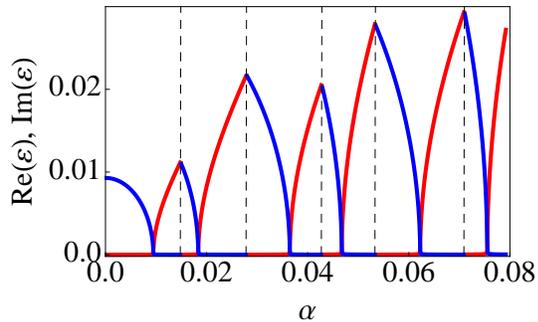}}
\end{center}
\caption{Plot of the absolute value of the real part (blue lines) 
and the imaginary part (red lines) of the 
lowest eigenvalue in the spectrum of $H_{\rm NL}$ for evanescent states with
angular momentum $m = 0$ in a 
semi-infinite cylindrical geometry with radius $R = 100 a$ 
($a$ being the microscopic length scale in the model), for 
$m_0 = 0.1$ eV, $m_1/a^2 = 2.0$ eV, $v/a = 1.0$ eV. The units  
are the same as in Fig. \ref{one}.}
\label{twob}
\end{figure}

Such a nontrivial realization of the invariance under $I$ and $T$ holds only in the 
case of the type A nodal line semimetal, as for type B all the evanescent states
have $k_z = 0$. In this latter case, they correspond to exceptional points in the 
spectrum of $H_{\rm NL}$ which fall along the imaginary axis in the $(k_z, \alpha )$ 
complex plane. A representation of the sequence of branch points for a 
type B nodal line semimetal with cylindrical geometry is shown in Fig. \ref{twob}.
We notice that the plot has a series of discontinuities, which arise from the
fact that the lowest eigenvalue of $H_{\rm NL}$ is found in different subbands as the 
value of $\alpha $ increases. In each continuous region, we observe a clear 
characterization of the branch point as the location where the real
and the imaginary part of the eigenvalue vanish.

The plot in Fig. \ref{twob} highlights an important property that applies
both to type A and type B nodal line semimetals. As already mentioned, it 
is clearly observed in the figure that exceptional points with 
different values of the complex momentum belong in general to different
subbands in the spectrum of $H_{\rm NL}$. This can be also appreciated (though less
neatly) in the representation of Fig. \ref{two}, where a careful inspection
shows that a line of discontinuity exists in the contour plot between each 
two consecutive branch cuts. From the point of view of the complex 
structure, this means that exceptional points with different values of 
$k_z + i \alpha $ belong in general to different branches of the Riemann 
sheet arising from the diagonalization of $H_{\rm NL}$ for complex momentum. 
This implies that it is not possible in general to undo the branch cuts 
by merging contiguous exceptional points. The only branch points that can 
be made to coalesce are those that pertain to the same branch in the complex 
structure  ---which, in the case of a type A semimetal, are those connected 
precisely by the $I$ and $T$ operations. This reassures once more the 
topological stability of the collection of drumhead surface states, 
implied in this framework by the underlying complex structure of the 
spectrum.

\section{Evanescent states in nodal line semimetal under electromagnetic radiation}
\label{sec:nodalfloquet}

A very interesting line of research is the control of quantum properties by external ac fields 
\cite{Dunlap86,Grossmann91,Holthaus95,Platero04,Martinez06,Creffield07}. In this search, 
it has been found that the effect of electromagnetic
radiation may change the properties of 2D semimetals, opening 
a gap in the bulk and leading to chiral currents at the boundary of the 
electron system\cite{oka,refael,inoue,demler,fu,arovas,GomezLeon13}. The effect of the 
radiation has been also investigated in the case of 3D Dirac and Weyl 
semimetals, finding that a circularly polarized photon field has the ability to shift 
the Dirac or Weyl points in momentum space\cite{Wang14,nara,palee}. 
New surface states have been also discovered in Weyl semimetals illuminated by monochromatic radiation, 
forming bands with macroscopic degeneracy and rotating currents\cite{Gonzalez16}. Another interesting result is that, using circularly polarized light in the proper direction, a nodal line semimetal can be transformed into a Weyl semimetal \cite{Yan16}.

In this section, we show that the idea of describing the surface states by 
exceptional points in the spectrum can be generalized to the case of the 
nodal line semimetals under electromagnetic radiation. This is a relevant
instance to check the topological protection that arises when extending
the momenta to the complex plane, since the effects of the radiation can 
be studied in regimes where it is not a small perturbation. 



The coupling to the vector potential can be done in the usual fashion,
adopting the Peierls prescription $\mathbf{k} \rightarrow \mathbf{k} 
+ \mathbf{A}$. In the case of circularly polarized radiation sent along the $z$ 
direction, we have
\begin{equation}
 \mathbf{A} = \left(A \cos (\Omega t), A \sin (\Omega t), 0 \right)
\end{equation}
The Hamiltonian becomes then
\begin{widetext}
\begin{eqnarray}
H  & = &   \left(m_0 + m_1 (\partial_x^2 + \partial_y^2 +
    2iA  \cos (\Omega t) \partial_x + 2iA \sin (\Omega t) \partial_y - A^2 +
        \partial_z^2 )  \right) \sigma_z - iv \partial_z \sigma_x          \\
   & = &  \left(m_0 + m_1 (\partial_x^2 + \partial_y^2 +
    iA e^{i\Omega t} \partial_- + iA e^{-i\Omega t} \partial_+ - A^2 +
        \partial_z^2 )  \right) \sigma_z - iv \partial_z \sigma_x 
\label{ham2}
\end{eqnarray}
\end{widetext}
with 
\begin{eqnarray}
\partial_-  & = & \partial_x - i\partial_y       \\
\partial_+  & = & \partial_x + i\partial_y 
\end{eqnarray}

We have the commutation rules with the angular momentum 
$L_z = -ix\partial_y + iy\partial_x$:
\begin{eqnarray}
\left[L_z , \partial_- \right] & = & - \partial_-       \\    
\left[L_z , \partial_+ \right] & = &  \partial_+ 
\end{eqnarray}
From these relations, it can be easily seen that 
\begin{eqnarray}
e^{iL_z \Omega t} \; \partial_- \; e^{-iL_z \Omega t} 
         & = & e^{-i \Omega t} \; \partial_-   \label{r1}   \\
e^{iL_z \Omega t} \; \partial_+ \; e^{-iL_z \Omega t} 
         & = & e^{i \Omega t} \; \partial_+
\label{r2}
\end{eqnarray}
We can now rely on Eqs. (\ref{r1})-(\ref{r2}) to pass to a time-independent
Hamiltonian by applying the unitary transformation 
\begin{equation}
U = e^{-iL_z \Omega t}
\end{equation}
We have
\begin{widetext}
\begin{equation}
\widetilde{H}   =   U^\dagger H U - i U^\dagger \partial_t U         
       =  \left(m_0 + m_1 (\partial_x^2 + \partial_y^2 +
    iA \partial_-  + iA \partial_+ - A^2 +
        \partial_z^2 )  \right) \sigma_z - iv \partial_z \sigma_x - \Omega L_z 
\label{hamt}
\end{equation}
\end{widetext}

We can now perform an analysis of the evanescent states under radiation, following
that accomplished before for the Hamiltonian $H_{\rm NL}$. The idea is to focus on
evanescent states decaying as
\begin{equation}
\chi (r, \theta, z)  \sim  e^{i k_z z} e^{-\alpha z} f(r,\theta)
\end{equation}
We look in particular for zero-energy modes, which must correspond to 
solutions of the equation
\begin{widetext}
\begin{equation}
\left[m_0 + m_1 \left( \frac{1}{r} \frac{\partial }{\partial r} 
  \left(r \frac{\partial }{\partial r} \right) + \frac{\partial_\theta^2}{r^2}  
      + iA \partial_-  + iA \partial_+ - A^2
       - k_z^2 + \alpha^2 - 2i k_z \alpha \right)  \right] \sigma_z \chi_{k_z}^{\mbox{}} 
      + v (k_z + i\alpha ) \sigma_x \chi_{k_z}^{\mbox{}}  
               + i\Omega \partial_\theta \chi_{k_z}^{\mbox{}} =  0
\label{eva}
\end{equation}
\end{widetext}

\begin{figure}[tb]
\begin{center}
\mbox{\epsfxsize 8.1cm \epsfbox{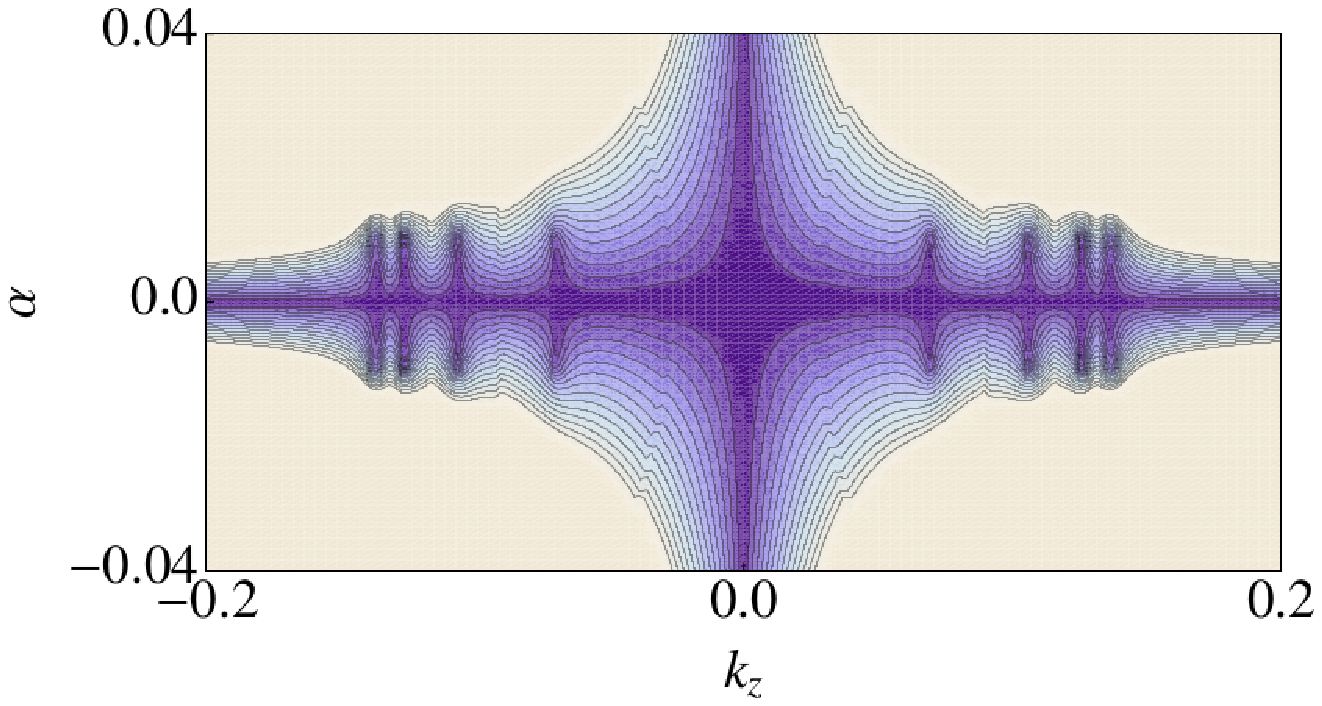}}
\end{center}
\begin{center}
(a)
\end{center}
\begin{center}
\mbox{\epsfxsize 8.1cm \epsfbox{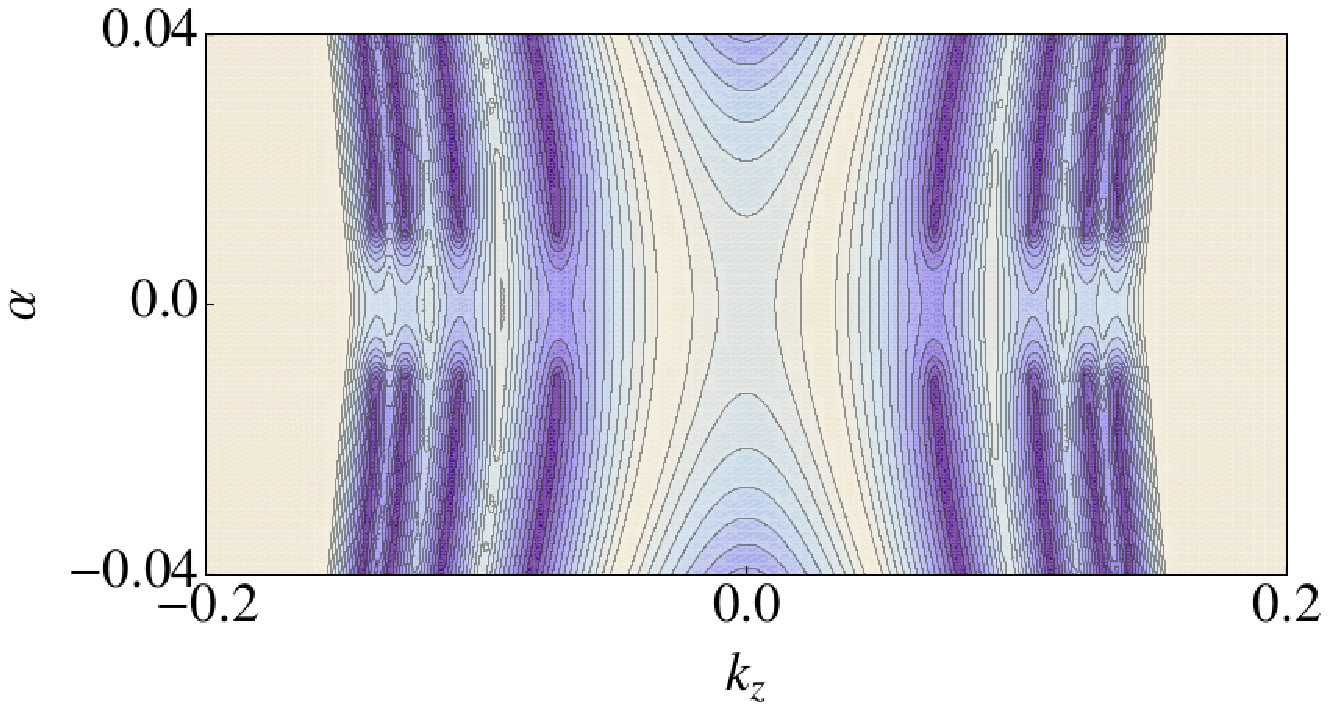}}
\end{center}
\begin{center}
(b)
\end{center}
\caption{Contour plots of the absolute value of the imaginary part (a) 
and the real part (b) of the 
lowest eigenvalue in the spectrum of $\widetilde{H}$ for evanescent states in a 
semi-infinite cylindrical geometry with radius $R = 100 a$ 
($a$ being the microscopic length scale in the model) 
for $m_0 = 0.01$ eV, $m_1/a^2 = 0.5$ eV, $v/a = 0.01$ eV, $Aa = 0.05$, $\Omega = 1.0$ eV. 
The units and intensity code are the same as in Fig. \ref{one}.}
\label{five}
\end{figure}

We perform first a numerical analysis of the problem, in which it is 
convenient to apply the gauge transformation 
\begin{equation}
\chi_{k_z}^{\mbox{}} = e^{-iAx} \; \widetilde{\chi}_{k_z}^{\mbox{}}
\end{equation}
This converts Eq. (\ref{eva}) into
\begin{widetext}
\begin{equation}
\left[m_0 + m_1 \left( \frac{1}{r} \frac{\partial }{\partial r} 
  \left(r \frac{\partial }{\partial r} \right) + \frac{\partial_\theta^2}{r^2}  
       - k_z^2 + \alpha^2 - 2i k_z \alpha \right)  \right] \sigma_z \widetilde{\chi}_{k_z}^{\mbox{}} 
      + v (k_z + i\alpha ) \sigma_x \widetilde{\chi}_{k_z}^{\mbox{}}  
               + i\Omega \partial_\theta \widetilde{\chi}_{k_z}^{\mbox{}} 
                - A \Omega \: r \sin (\theta ) \widetilde{\chi}_{k_z}^{\mbox{}} =  0
\label{eva2}
\end{equation}
\end{widetext}
Then one can check that it is possible to adjust the values of $k_z$ and 
$\alpha $ to obtain solutions of Eq. (\ref{eva2}). The numerical resolution can 
be done for instance in a cylinder with $r < R$, in which there is a finite 
number of solutions depending on the radius $R$. These can be more easily 
visualized computing the spectrum of the operator in Eq. (\ref{eva2}) in the 
complex plane $(k_z , \alpha )$, which leads in general to a picture like that
represented in Fig. \ref{five}.

We observe that the zero-energy modes correspond once again to exceptional
points in a spectrum of complex eigenvalues, with a structure similar to that
already found in the absence of electromagnetic radiation. The exceptional
points are easily identified as branch points with a tail where the lowest 
eigenvalue has zero imaginary part, and an opposite tail where the real 
part of the eigenvalue vanishes. This is the typical behavior for a square root 
singularity, which is also consistent with the structure of the branch cuts 
connecting homologous branch points across the $k_z$ axis, as seen in 
Fig. \ref{five}.

We can complement the numerical approach with an analytic search of the 
solutions of Eq. (\ref{eva}) when the surface has infinite size. We can 
start with the set of states spanned by the basis
\begin{eqnarray}
\chi_{k_z,+}^{(m)} (r, \theta , z)  & = &   e^{i k_z z} e^{-\alpha z} 
               e^{im \theta } J_m (\hat{k}_r r )  | + \rangle       \\
\chi_{k_z,-}^{(m)} (r, \theta , z)  & = &   e^{i k_z z} e^{-\alpha z} 
               e^{im \theta } J_m (\hat{k}_r r )  | - \rangle 
\end{eqnarray}
where 
\begin{equation}
\hat{k}_r = \sqrt{\tfrac{m_0}{m_1} - k_z^2 + \alpha^2 - A^2 - \delta^2 }
\label{hkr}
\end{equation}
In this case, we keep $\alpha $ and $\delta $ as free parameters that we have 
to adjust in order to find the solutions of Eq. (\ref{eva}). This provides us 
with a very flexible collection of states, which proves to be large enough to
capture the evanescent zero-energy modes. 

The action of the off-diagonal perturbations induced in (\ref{eva}) by the 
radiation is given by 
\begin{eqnarray}
 \partial_- \sigma_z \chi_{k_z,\pm}^{(m)}  & = &   \hat{k}_r  \chi_{k_z,\mp}^{(m-1)}   \\
 \partial_+ \sigma_z \chi_{k_z,\pm}^{(m)}  & = & - \hat{k}_r  \chi_{k_z,\mp}^{(m+1)}
\end{eqnarray}
Then, the states we obtain by the action of $\widetilde{H}$ remain in 
the subspace we have defined:
\begin{widetext}
\begin{eqnarray}
\widetilde{H} \chi_{k_z,+}^{(m)}   & = &  im_1 A \hat{k}_r \chi_{k_z,-}^{(m-1)} 
  - im_1 A \hat{k}_r \chi_{k_z,-}^{(m+1)}  + m_1 \delta^2 \chi_{k_z,-}^{(m)}  
   - \left( v \alpha + i(2m_1 \alpha - v) k_z \right) \chi_{k_z,-}^{(m)}
                                            -  m \Omega  \chi_{k_z,+}^{(m)}      \\
\widetilde{H} \chi_{k_z,-}^{(m)}   & = &  im_1 A \hat{k}_r \chi_{k_z,+}^{(m-1)} 
  - im_1 A \hat{k}_r \chi_{k_z,+}^{(m+1)}  + m_1 \delta^2 \chi_{k_z,+}^{(m)} 
   - \left( - v \alpha + i(2m_1 \alpha + v) k_z \right) \chi_{k_z,+}^{(m)}  
                                        -  m \Omega  \chi_{k_z,-}^{(m)}   
\end{eqnarray}
\end{widetext}

\begin{figure}[h]
\begin{center}
\includegraphics[height=3.5cm]{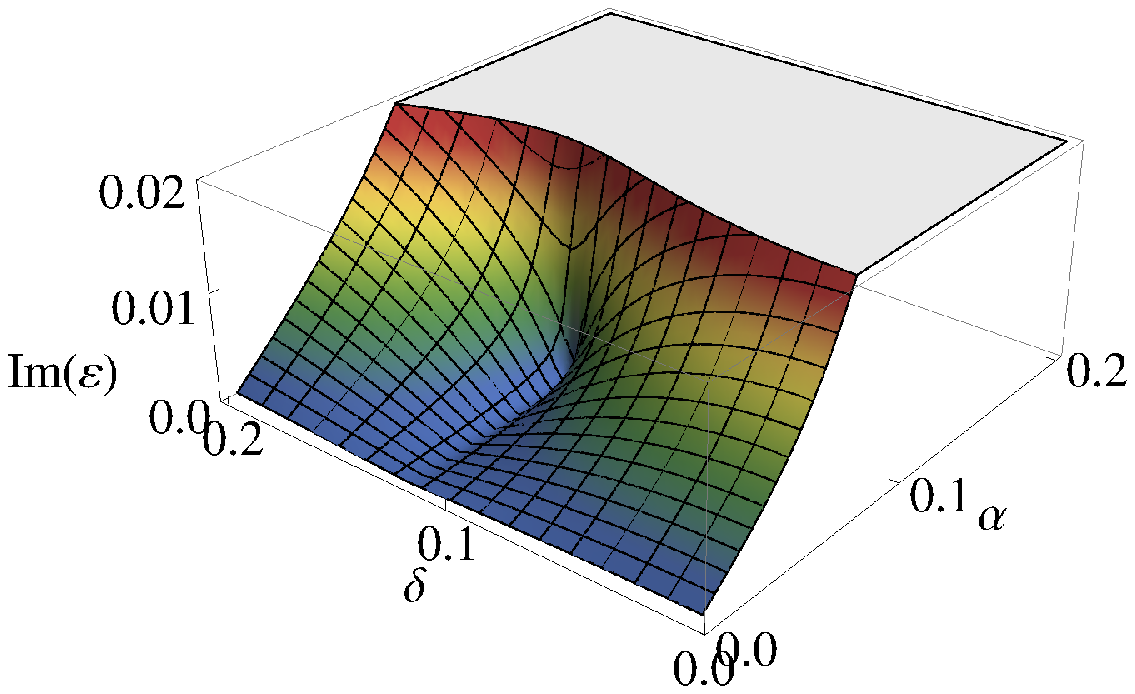}
\hspace{9.0cm}  (a)\\
\mbox{}\\
\includegraphics[height=3.0cm]{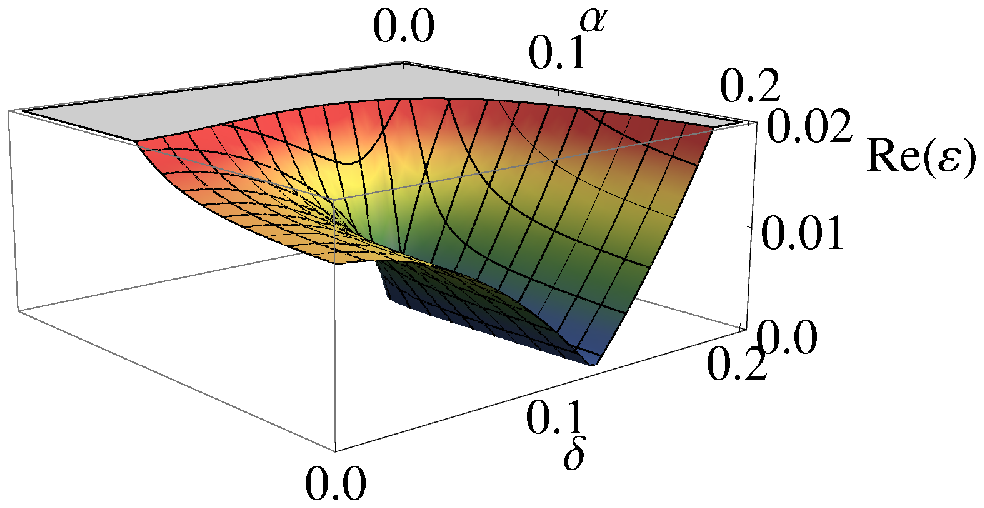}
 \hspace{5.5cm} (b)
\end{center}
\caption{Plot of the absolute value of the imaginary part (a) 
and the real part (b) of the lowest 
eigenvalue of $\widetilde{H}$ for evanescent states with variable
parameters $\alpha $ and $\delta $, for $k_z a = 0.2$ and 
$m_0 = 1.0$ eV, $m_1/a^2 = 1.0$ eV, $v/a = 0.2$ eV, $Aa = 0.1$, $\Omega = 0.5$ eV. 
The units are the same as in Fig. \ref{one}.}
\label{seven}
\end{figure}

One can check that, for not too large values of $k_z$ and $A$, it is possible to find 
a point in the space of parameters $(\alpha , \delta )$ where the eigenvalue 
of $\widetilde{H}$ becomes zero. This arises as a branch point in the spectrum 
of complex eigenvalues, as represented in Fig. \ref{seven}. 
This procedure works up to certain limit values of $A$ and $k_z$, beyond which 
the branch point is lost and there is no signature of zero-energy modes.

\begin{figure}[h]
\begin{center}
\includegraphics[height=5.5cm]{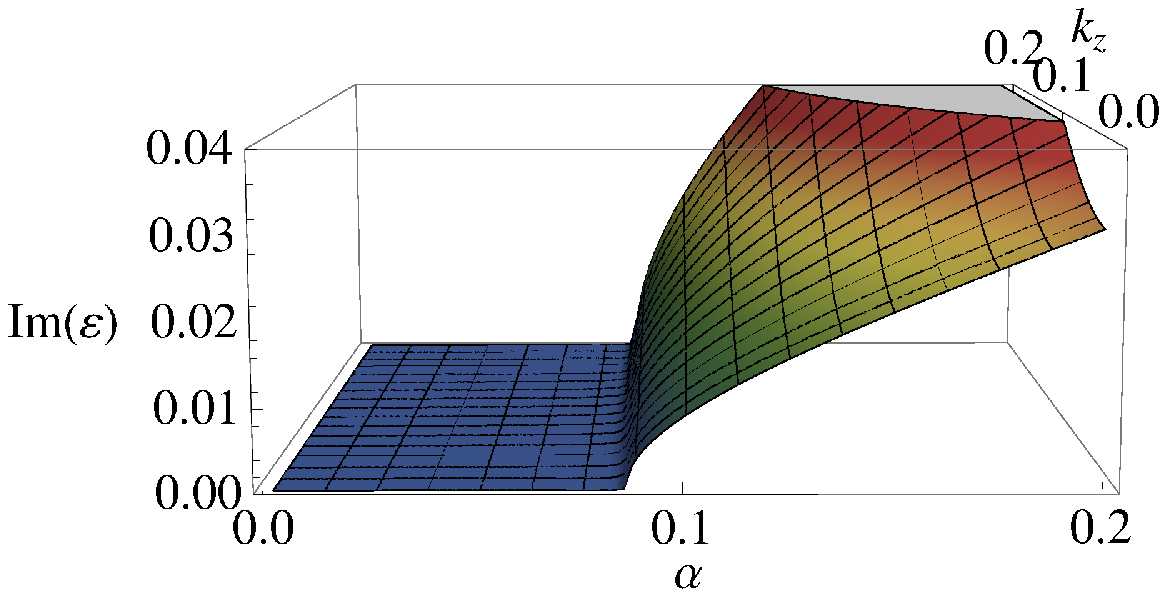}
 \mbox{} \hspace{9.0cm}  (a)\\
\includegraphics[height=5.5cm]{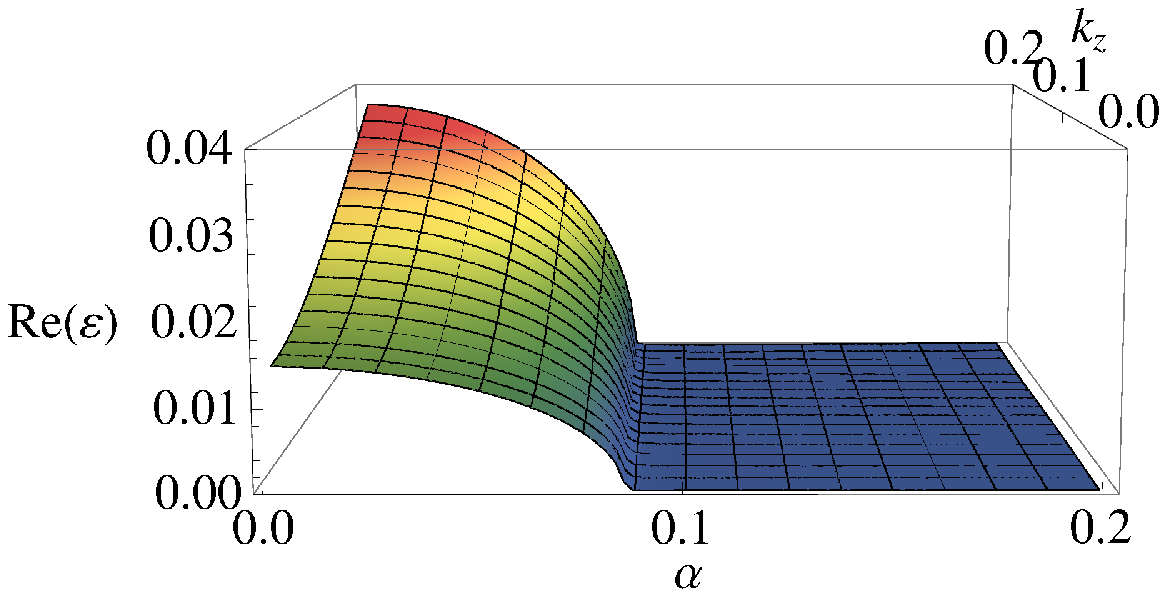}
 \hspace{5.5cm} (b)
\end{center}
\caption{Plot of the absolute value of the imaginary part (a) 
and the real part (b) of the lowest 
eigenvalue of $\widetilde{H}$ for evanescent states with complex 
momentum $k_z + i\alpha $, for $m_0 = 1.0$ eV, $m_1/a^2 = 1.0$ eV, $v/a = 0.2$ eV, $Aa = 0.1$, 
$\Omega = 0.5$ eV. The units are the same as in Fig. \ref{one}.}
\label{eight}
\end{figure}

If we plot the spectrum as a function of $k_z + i\alpha $, once the appropriate 
value of $\delta $ is set, we get the image shown in Fig. 
\ref{eight}. We observe there a line of zero-energy modes, which may be thought 
as the accumulation of the exceptional points already found in the finite-size 
numerical approach. This shows that a continuum of evanescent modes persist under 
the effect of the electromagnetic radiation, arising as a result of the 
hybridization of drumhead states with different angular momentum and being 
labeled by the component $k_z$ of the momentum.

We have to remark however that, in the present case, the exceptional points
we have found do not imply the existence of zero-energy surface states attached
to a given boundary of the nodal line semimetal. This is so as the evanescent 
eigenstates of $\widetilde{H}$ do not realize the symmetry required to build
appropriate linear combinations that may guarantee the vanishing of the surface 
states at the boundary, say at $z = 0$. In the case of the unperturbed nodal 
line semimetal, we saw above that the crucial symmetry for the existence of 
zero-energy surface states was given by the $T$ operation in (\ref{tr}). The introduction 
of the radiation field breaks the time-reversal invariance, which explains 
that the evanescent modes corresponding to momenta $k_z$ and $- k_z$ have 
different wave functions at $z = 0$ in the presence of the radiation. This stresses the significance of the 
symmetry of the Hamiltonian to guarantee the stability of the zero-energy surface states,
together with the topological protection already provided by the existence of 
the exceptional points in the spectrum.

\section{Conclusions}
\label{sec:conclusions}

We have shown how the surface states in topological semimetals are associated to exceptional points of the spectrum
in the extension of the Brillouin zone to complex values of momenta. These exceptional points are
very robust under perturbations of the Hamiltonian as they can be only annihilated by merging them in pairs to close the branch cuts in the spectrum. This draws a useful way of understanding the topological protection of evanescent states like those forming the Fermi arcs in Weyl semimetals or the drumhead states in nodal line semimetals. In this regard, the mechanism of topological protection seems to be rather different to that of the nodes of the semimetals themselves (and different also to that studied recently in the case of 3D Dirac semimetals \cite{Kargarian16}). 

We have also shown that the evanescent states decay exponentially into the semimetal, and we have illustrated the dependence of the penetration length according to the parameters of a model Hamiltonian for both Weyl and nodal line semimetals. The penetration into the semimetal can be very short and oscillating, or longer and with pure exponential decay, depending on the ratio between the linear and the quadratic terms in the Hamiltonian. According to this ratio, we have classified the topological semimetals as type A or type B. Although we expect the type A (the short penetration with oscillations) to be realized in real materials, the proposals for artificial topological semimetals in photonic or cold atom setups could easily reach both regimes \cite{Dubcek15,Xu2,Lu13,Lu15}. The distinction between type A and type B is also formally related to the different way in which the branch cuts corresponding to the exceptional points are arranged in the complex plane. 

In nodal line semimetals, we have also studied the effect of an external ac field on the surface states. Evanescent states with different angular momentum are mixed by the external field. Interestingly, the resulting states are still in correspondence to exceptional points in the complex plane and they are protected from small perturbations as such. These mixtures of states with different angular momenta should carry a rotating current similarly to that studied before in Weyl semimetals \cite{Gonzalez16}. However, the reduction of symmetry from the ac field implies that the zero-energy surface states of nodal line semimetals are in general not stable in the presence of radiation, 
as this breaks the required invariance to comply with appropriate boundary conditions.  

We believe our work paves the way for an alternative understanding of topological protection of edge states in gapless systems, based on the extension of the band structure for complex values of the momenta. In this complex momentum space, the surface states arise as exceptional points, making possible to study their stability by discerning the perturbations capable to close the branch cuts in the complex band structure. It would be an interesting future avenue to investigate the connection between this description of surface states and
the topological properties of photonic materials or open quantum systems that can be described by actual non-Hermitian Hamiltonians \cite{Esaki11,Diehl11,Zeuner15,SanJose16,Peng16}.

\acknowledgments

We acknowledge financial support through Spanish grants MINECO/FEDER No. FIS2015-63770-P and No. FIS2014-57432-P.

\end{document}